\documentclass[twocolumn,showpacs,preprintnumbers,amsmath,amssymb]{revtex4}

\usepackage{graphicx}
\usepackage{dcolumn}
\usepackage{bm}

\begin{document}


\title{Effect of scattering and contacts on current and 
electrostatics in carbon nanotubes}

\author{A. Svizhenko
}\email{svizhenk@nas.nasa.gov}
 \affiliation{Center for Nanotechnology and NASA 
 Advanced Supercomputing Division,\\
NASA Ames Research Center,\\
Mail Stop: 229-1               
Moffett Field, CA 94035-1000 \\} 

\author{M. P. Anantram}
\email{anant@mail.arc.nasa.gov}
\affiliation{Center for Nanotechnology and NASA 
Advanced Supercomputing Division,\\
NASA Ames Research Center,\\
Mail Stop: 229-1
Moffett Field, CA 94035-1000 \\}%

\date{\today}

\begin{abstract}
We computationally study the electrostatic 
potential profile and current carrying capacity 
of carbon nanotubes as a function of length and diameter. 
Our study is based on solving the non equilibrium 
Green's function and Poisson  equations self-consistently, 
including the effect of electron-phonon scattering. 
A transition from ballistic to diffusive regime of electron transport 
with increase of applied bias
is  manifested by qualitative changes in potential profiles, 
differential conductance and electric field in a nanotube. 
In the low bias ballistic 
limit, most of the applied voltage drop occurs near the contacts. 
In addition, the electric field at the tube center 
increases proportionally  with diameter. 
In contrast, at high biases, most of the applied voltage 
drops across the nanotube, and the electric field at the tube center 
decreases with increase in diameter. 
We find that the differential conductance can
increase or decrease 
with bias as a result of an interplay 
of nanotube length, diameter
and a quality factor of the 
contacts. From an application view point, we find that the current 
carrying capacity of nanotubes increases with increase in diameter. 
Finally, we investigate the 
role of inner tubes in affecting the current carried by the 
outermost tube of a multiwalled nanotube.
\end{abstract}



\maketitle

\section{\label{sec:intro}Introduction}
Metallic carbon nanotubes are near ideal conductors of 
current~\cite{frank-science-98,pablo-apl-99,poncharal-epjd-99,
nygard-applphysa-99,kong-prl-01}. While a single nanotube can be 
used as an interconnect in molecular electronics, nanotube arrays 
have shown promise as more conventional interconnects in 
conjunction with silicon technology~\cite{kreupl-microeleceng-02,
ngo-eeetn-04}. A single nanotube has two modes that carry current 
at the Fermi energy, which yields a low bias conductance of 
$4e^2/h$ and resistance of 6.5 $k \Omega$. This corresponds 
to a current of 155 $\mu$A at a bias of 1V. Noting that a 
nanotube diameter can be as small as 5 $\AA$, it is easy to 
estimate that an array of metallic carbon nanotubes can carry 
current densities larger than $10^{10}\;A/cm^{2}$. In fact, 
current densities approaching $10^9 \; A/cm^2$ have been 
demonstrated~\cite{pablo-apl-99,wei-apl-01}. 

The diameter of both metallic and semiconducting nanotubes can vary from 
5 $\AA$ to many tens of nanometers, with the electronic properties 
determined by the chiral angle. The bandgap of semiconducting nanotubes 
decreases inversely with diameter. As a result, a large diameter nanotube
 with a radius of 19 nm will have a bandgap of less than $2.5\;kT$ at 
 room temperature, meaning that large diameter semiconducting nanotubes 
 carry non negligible current. Further, electrical contact can be made 
 to many shells of large diameter multiwalled carbon nanotubes.  
 Reference \cite{pablo-apl-99} demonstrated a resistance of 
 nearly 500 $\Omega$ in a multiwalled sample, which corresponds to 
 about twelve conducting shells. Therefore, both small and large 
 diameter nanotubes are promising as interconnects. 
 Experiments on small diameter carbon nanotubes show that the differential 
conductance decreases with applied bias, for voltages larger than 
150 mV~\cite{yao-prl-00,collins-prl-01,pablo-prl-02}.
Reference \cite{yao-prl-00} showed that the conductance decreases with 
increase in bias was caused by reflection of electrons incident in 
crossing subbands due to scattering with zone boundary phonons. The 
non crossing subbands of small diameter nanotubes do not carry current 
due to their large band gap. In contrast, large diameter nanotubes 
experimentally show an increase in conductance with applied 
bias~\cite{frank-science-98,poncharal-epjd-99,poncharal-jpcb-02,
liang-apl-04}. 
Reference \cite{Anantram-prb-00} suggested that as the diameter 
increases, electrons may tunnel into non crossing subbands, thus 
causing an increase in differential conductance with applied bias. 
The main drawbacks of the calculation in reference \cite{Anantram-prb-00} 
was that the results depended on the assumed form of potential drop 
in metallic nanotubes and further, electron-phonon scattering was 
neglected. Recently, we performed self-consistent calculations 
\cite{svizhenko-cnt-ball}, which showed a dramatic increase in 
differential conductance of large diameter nanotubes with bias, 
in the ballistic limit. In the current work, we present results 
for current flow and potential profile in metallic nanotubes 
from a more comprehensive model, which includes both {\it charge 
self consistency} and {\it electron-phonon scattering}. The 
potential and current-voltage characteristics are studied as a 
function of nanotube diameter and length. We are primarily 
interested in short ($\sim$ 100 nm) rather than long nanotubes, 
where the physics is more interesting and technological applications 
are promising.
The nature of the metallic contacts is also
important. From an experimental view point the contact between a metal and 
a nanotube can either be an end-contact or side-contact. 
The end-contact corresponds to only the nanotube tip electronically 
interacting with the metal contact. In experiments, the end-contacts 
usually involves strong chemical modification of the nanotube at the 
metal-nanotube interface~\cite{kong-prl-01}. Also, reference 
\cite{frank-science-98} found that end-contacts without sufficient 
chemical modification of the nanotube-metal interface have a large 
contact resistance.
Due to the uncertainty of the contact bandstructure, modeling experimental 
end-contacts even remotely correct is difficult.
      The side-contacts correspond to 
   coupling between metal and nanotube atoms over many unit cells
    of the nanotube, and can be thought of as a nanotube buried 
    inside a metal. Most experimental configurations correspond to
     side-contacts \cite{frank-science-98,tans}. 
     An important feature of the side contact is that the  coupling 
     between
      atoms in the nanotube is much stronger than coupling between 
      nanotube and metal atoms, which means that the bandstructure 
      of the contacts is similar to that of the nanotube between 
      the contacts.     
     In the side-contact geometry, 
      electrons are predominantly injected from the metal into the
       nanotube buried in metal and then transmitted to the 
       nanotube region between contacts.              
        In fact, as proof of such a process, scaling 
       of conductance with contact area has been observed in the 
       side-contacted geometry by references \cite{frank-science-98,tans}. 
       Modeling has also shown that the conductance in metallic zigzag 
       nanotubes can be close to the theoretical maximum of $4e^2/h$, 
       when there is
       sufficient overlap with the 
       contact \cite{Anantram-apl-01}. In this work, we consider the 
       metal-nanotube contact in both the limiting cases of side- and 
       end-contacts.

The outline of the paper is as follows. In section \ref{sec:formalism},
 we describe the formalism. The electrostatics at low bias is presented 
 in section \ref{sec:low}, electrostatics at high bias and current-voltage 
 characteristics are presented in section \ref{sec:high}. The role of 
 inner shells in affecting the potential profile of a current carrying 
 outer shell is described in section \ref{sec:shells}. End-contacts 
 that form both good and poor contacts are discussed in section 
 \ref{sec:contacts}. We present our conclusions in section \ref{sec:conc}.

\section{\label{sec:formalism} Formalism}
In this paper we consider only zigzag carbon nanotubes. 
The analysis for armchair nanotubes is similar.
The general form of the Hamiltonian for electrons in a carbon nanotube
 can be written as:
\begin{eqnarray}
H & = & \sum_{i,x}U^{i}_{x}c^{\dagger}_{i,x}c_{i,x}+
\sum_{i,j,x,x'}
t^{i,j}_{x,x'}c^{\dagger}_{i,x}c_{j,x'} 
\end{eqnarray}
The sum is taken over all rings $i,j$ in transport direction and all atoms located at $x,x'$ 
in each ring.

\begin{figure}
\includegraphics [width=9.0cm] {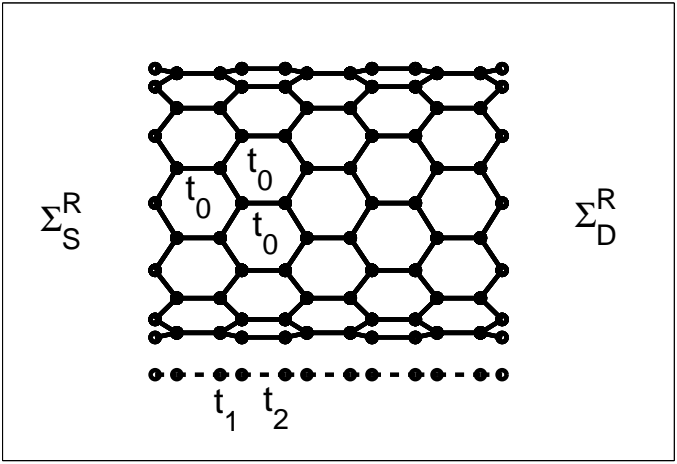}
\caption{Zigzag carbon nanotube and the corresponding 1D chain. 
The hopping parameter between nearest neighbors in the nanotube is $t_o$. 
The 1-D 
chain has two sites per unit cell
with on-site potential $U_o$ and 
hopping parameters $t_1=2t_o
\cos({\tilde qa \over 2})$ and $t_2=t_o$ }
\label{fig:tube}
\end{figure}

We make the following common approximations:
i)  only nearest neighbors are included; 
each atom in an $sp^2$-coordinated carbon nanotube
has three nearest neighbors, located $a_{cc}=$ 1.42 $\AA$ away;
ii) the bandstructure consists of only $\pi$-orbital, with the 
hopping parameter
$t_o=V_{pp\pi}=-2.77$ eV and the on-site potential $U_o=\epsilon_p=0$. 
Such a tight-binding model 
is adequate to model transport properties in undeformed nanotubes.
Within these approximations, only the following parameters are non zero:
\begin{eqnarray}
U^{i}_{x}& = & U_o , \forall i \nonumber \\
t^{i,i-1}_{x,x'}& = &t^{i-1,i}_{x,x'}=t_o\delta_{x\pm a/2,x'}, \forall i=2k 
\label{eq:hop_par}  \\
t^{i,i+1}_{x,x'}& = &t^{i+1,i}_{x,x'}=t_o\delta_{x,x'}, \forall i=2k ~,
\nonumber
\end{eqnarray}
where $a=a_{cc}\sqrt{3}$.
 In $(N,0)$ zigzag nanotubes, the 
wave vector in the circumferential direction is quantized as 
$\tilde q=2\pi q/Na$, $q=1,2,...N$, 
creating eigenmodes in the energy spectrum. 
By doing a Fourier expansion of $c^{\dagger}_{i,x}$ and $c_{i,x}$ 
in $\tilde q$-space  and using  Eq.~(\ref{eq:hop_par}) we obtain
a decoupled electron Hamiltonian in the eigenmode space:
\begin{eqnarray}
H & = & \sum_{q} H^{q}\\
H^{q}& = & \sum_{i}
(U^{i}_{\tilde q}
c^{\dagger}_{i,\tilde q}c_{i,\tilde q}
+t^{i,i\pm1}_{\tilde q}
c^{\dagger}_{i,\tilde q}c_{i\pm1,\tilde q})
\end{eqnarray}
where
\begin{eqnarray}
U^i_{\tilde q} &=& U_o, \forall i \nonumber\\
t^{i,i-1}_{\tilde q} & = & t^{i-1,i}_{\tilde q}=2t_o
\cos({\tilde qa \over 2}) \equiv t_1, \forall i=2k  \\
t^{i,i+1}_{\tilde q}& = & t^{i+1,i}_{\tilde q}=
t_o \equiv t_2,  \forall i=2k  \nonumber
\end{eqnarray}
The 1-D tight-binding Hamiltonian $H^q$ describes  a  
chain with two sites per unit cell
with on-site potential $U_o$ and 
hopping parameters $t_1$ and $t_2$ (Fig.\ref{fig:tube}).
For numerical solution, the spatial grid corresponds to the rings of the nanotube, 
separated by $a/2$ with a unit cell 
length of $3a/2=$ 2.13 $\AA$, which is half the unit cell length of a 
zigzag nanotube. The number of spatial gridpoints $Ny$ is 
equal to the number of rings.

The subband dispersion relations are given by
\begin{eqnarray}
E_q(k)=\pm |t_o|{\bigl| 1+4cos({3ak\over2})cos({q\pi\over N})+
4cos^2({q\pi\over N})\bigr|}^{1/2}
\end{eqnarray}
Therefore, when $N=3k$,
 there are two subbands with zero bandgap 
 the tube is metallic. In the rest of the paper we  
 distinguish between metallic 
or {\it crossing} subbands  ($q=N/3$ and $2N/3$) and semiconducting or 
{\it non crossing} subbands.

For each subband $q$ we solve a system of transport equations
 \cite{JAP}:
\begin{eqnarray}
AG^{R,q}&=&I \label{eq:tran1}\\
AG^{<,>,q}&=&(\Sigma_{c,q}^{<,>}(E)+\Sigma^{<,>}_{ph,q}(E))G^{A,q}~, 
\label{eq:tran2}
 \end{eqnarray}
where $A=E-H^q-V-\Sigma_{c,q}^{R}(E)-\Sigma^R_{ph,q}(E)$. The self-energies
$\Sigma_{c,q}^{R,<,>}(E)$ and $\Sigma^{R,<,>}_{ph,q}(E)$ 
represent the effect of contacts and scattering.

 The contacts 
are assumed to be reflectionless reservoirs
 maintained at equilibrium, i.e. they have  well defined chemical 
potentials, equal to that of the metal leads: 
$V_S$ in the source and $V_D$  in 
the drain.
Further, the nanotube and metal
 are assumed to have the same workfunction.
The contact self-energies $\Sigma_{c,q}^{R,<,>}=
\Sigma_{S,q}^{R,<,>}+\Sigma_{D,q}^{R,<,>}$ due 
to the source contact are found
\cite{JAP} using
the surface Green's function $g^{S}_q$ of the leads, 
which is the solution 
of the following system of equations:
\begin{eqnarray}
(a_1-t_{2}^2g^{S}_{2,q}-\Sigma^R_{ph,1,q})g^{S}_{1,q}&=&1 \nonumber\\
(a_2-t_{1}^2g^{S}_{1,q}-\Sigma^R_{ph,2,q})g^{S}_{2,q}&=&1 ~, \label{eq:surf_gr}
\end{eqnarray}
where the indices $1$ and $2$ stand for the two sites of the unit cell,
$a_{1,2}=E-U_o-V_{S}$ and $\Sigma^R_{ph,1,2,q}$ are electron-phonon 
self-energies at the first two nodes near the source.
The Green's function for the drain
are solved for in a similar way, by making the  substitutions
$t_{1,2}^{S} \rightarrow t_{2,1}^{D}$, 
$g_{1,2,q}^{S}\rightarrow g_{1,2,q}^{D}$ and taking 
$\Sigma^R_{ph,1,2,q}$ at the drain end. The expression for 
contact self-energies is then given by:
\begin{eqnarray}
\Sigma^R_{S,q}&=&t_{2}^2g^{S}_{2,q} \nonumber \\
\Sigma^R_{D,q}&=&t_{1}^2g^{D}_{1,q} \nonumber \\
\Sigma^{<}_{S,q}&=&-2if_S\Im m[\Sigma^R_{S,q}] \nonumber \\
\Sigma^{<}_{D,q}&=&-2if_D\Im m[\Sigma^R_{D,q}]  \label{eq:surf_sigma} \\
\Sigma^{>}_{S,q}&=&2i(1-f_S)\Im m[\Sigma^R_{S,q}] \nonumber \\
\Sigma^{>}_{D,q}&=&2i(1-f_D)\Im m[\Sigma^R_{D,q}]~,\nonumber 
\end{eqnarray}
where $f_{S,D}$ are the Fermi factors in the source and drain leads.

The electron-phonon scattering is treated within self-consistent 
Born approximation. 
The electron-phonon  self-energies are due to elastic 
(acoustic phonon) and inelastic (optical and zone-boundary phonon)
 scattering:
\begin{eqnarray}
\Sigma_{ph,q}^{R,<,>}=\Sigma_{el,q}^{R,<,>}+
\Sigma_{in,q}^{R,<,>}~, \label{eq:phon1}
\end{eqnarray}
\begin{eqnarray}
\Sigma_{el,q}^{<,>}(E)&=&\sum_{q'}D_{el,q,q'}
G^{<,>,q'}
 \label{eq:phon2}~, \\
\Sigma_{in,q}^{<}(E)&=&\sum_{\nu,q'}D_{in,\nu, q,q'} \nonumber\\
&&[(n_B(\hbar\omega_{\nu})+1)G^{<,q'}
(E+\hbar\omega_{\nu})\nonumber\\
&&+n_B(\hbar\omega_{\nu}) G^{<,q'}
(E-\hbar\omega_{\nu})]~, \label{eq:phon3}\\
\Sigma_{in,q}^{>}(E)&=&\sum_{\nu,q'}D_{in,\nu, q,q'}\nonumber\\
&&[(n_B(\hbar\omega_{\nu})+1)G^{>,q'}
(E-\hbar\omega_{\nu})\nonumber\\
&&+n_B(\hbar\omega_{\nu}) G^{>,q'}
(E+\hbar\omega_{\nu})]~, \label{eq:phon4}\\
\Im m[\Sigma_{ph,q}^{R}(E)]&=&
[\Sigma_{ph,q}^{>}-\Sigma_{ph,q}^{<}]
/2i
~.\label{eq:phon5}\\
\Re e[\Sigma_{ph,q}^{R}(E)]&=&
{1 \over \pi} {\bf P}\int_{-\infty}^{+\infty} 
{\Im m[\Sigma_{ph,q}^{R}(E')] \over E'-E}dE'
\label{eq:kram}
\end{eqnarray}

The matrix elements squared due to particular scattering mechanisms
are chosen so as to satisfy 
experimentally  measured values  of the mean free path 
$\lambda_{o}^{el,in}$. Reference \cite{park-nl-04} reported 
$\lambda_{o}^{el} \approx$ 1.6 $\mu m$ and 
$\lambda_{o}^{in}\approx$ 10 nm for a tube 
with a diameter of 1.8 nm, corresponding to a (24,0) nanotube.
Inelastic scattering is due to zone boundary and optical 
phonon modes with energies of 160 and 200 meV.
Since, matrix elements squared scale inversely with the chirality index 
\cite{golgsman-prb-03},
one obtains for an $(N,0)$ nanotube 
\begin{eqnarray}
D_{el,in} \approx {\hbar\over 2\pi}{ 1 \over DOS(E_F) \cdot \tau_{el,in}} = 
 {at^2_o \over 2\lambda_{o}^{el,in}}({24 \over N})
~. \label{eq:scatt}
\end{eqnarray}
The decrease of a scattering rate in a single subband with 
the chirality index is saturated by the eventual increase of 
the intersubband scattering due to the larger number of subbands. 
Thus, with the increase of the diameter, the mean free path 
of a nanotube  approaches that of graphite.

Electron charge  and current density $n_i$ and $J_i$  at each node $i$ 
are found from the following equations:
\begin{eqnarray}
n_i&=&-{2i}\sum_{q} 
\int_{V_{min}}^{-eV_S+10kT}G^{<,q}_{i,i}(E){dE \over 2\pi} \label{eq:dens} \\ 
J_i&=&{4e \over \hbar }\sum_{q} \int_{-eV_D-10kT}^{-eV_S+10kT}
G^{<,q}_{i,i+1}(E)t^q_{i+1,i}{dE \over 2\pi}.  \label{eq:curr}
\end{eqnarray} 
The lower limit of integration of Eq.~(\ref{eq:dens}) is determined by 
$V_{min}=-3t_o-V_D-\Delta E_{rn}$,
 where $\Delta E_{rn}$ is a renormalization energy 
related to the real part of electron-phonon self-energy.

We model the electrostatics of the nanotube as a system of 
point charges  between the two contacts located at $y=y_S=0$ and 
$y=y_D=L$.
The "perfect contacts" are modeled as parallel semi-infinite 
three dimensional
 metal leads that are maintained at fixed source and drain potentials: 
 $V_S$ for $y<y_S$ and $V_D$ for $y>y_D$. 
 So, while the self-energies due to contacts are identical to that 
 of a semi-infinite nanotube, the role of electrostatics is included
  by image charges corresponding to a perfect metal. 
  The electrostatic potential consists of  a linear drop due to a 
uniform electric field
created by the leads and the potential due to the charges on the 
tube and their images
\begin{eqnarray}
V=-eV_S-e(V_D-V_S)(y-y_S)/(y_D-y_S)\nonumber \\
+\sum_{j}G(i,j)(n_j-N) \label{eq:prof}
\end{eqnarray}
with the Green's function  
{\small \begin{eqnarray}
G(i,j)&=&{e\over 4N\pi\epsilon_o}\sum_{n=-\infty}^{+\infty} \sum_{l}
\biggl[ {1\over\sqrt{(y_i-y_j+2nL)^2+\rho_{k,l}^2}} \nonumber\\
& & \;\;\;\;\;\;\;\;\;\;\;\;\;\;\;\;
-{1\over\sqrt{(y_i+y_j+2nL)^2+\rho_{k,l}^2}}\biggr]~.  \label{eq:green}
\end{eqnarray}}
Here,  $\rho_{k,l}$ is the radial projection of the vector between atom 
$k$ at ring $i$ and atom $l$ at ring $j$. The summation is 
performed over all atoms $l$ at ring $j$ for an arbitrary value of $k$.
Maintaining 
the nanotube atoms buried in the metal at a fixed potential is close 
to reality because of the screening properties of 3D metals. Within 
a few atomic layers from the metal surface, the potential should have
 approached the bulk values. While the variation in potential in these
  few atomic layers of the 3D metal is not captured in our model, our
   conclusions on the nanotube electrostatics should not be significantly 
   affected. 

The calculations for one bias point involve 
two simultaneous iterative processes: 
Born iterations for electron-phonon 
self-energies [Eqs.~(\ref{eq:tran1}-\ref{eq:kram})]
and Poisson iterations of [Eqs.~(\ref{eq:tran1}-\ref{eq:prof})] 
for the potential profile and charge distribution.   
Typically, $3-5$ Born iterations 
were performed before updating a potential profile using Eq.~(\ref{eq:prof}).
The solution of Eqs.~(\ref{eq:tran1}-\ref{eq:tran2}) employs 
a recursive algorithm \cite{JAP}, which scales linearly  
with the number of nodes. 

\subsection{\label{sec:numerics} The importance of 
self-consistency and proper treatment of electron-phonon scattering }
 A major computational burden of the approach discussed above are the
self-consistent procedure itself and also 
the Kramers-Kr\"onig relation [Eq.~(\ref{eq:kram})].
In order to take the integral in Eq.~(\ref{eq:kram}), at first,
we have to solve Eqs.~(\ref{eq:tran1}-\ref{eq:phon5}) 
over the whole band $[-3t_o-V_D-\Delta E_{rn}; +3t_o+\Delta E_{rn}]$.
Due to the presence of inelastic scattering, 
the energy grid has to be uniform. Typically, $Ne=10^4$ energy grid 
points are used 
in order to achieve a required precision in computing charge.
At second, Eq.~(\ref{eq:kram}) requires $Ne^2Ny$ 
multiplications for each subband, which makes it very time consuming.
Such computational requirements pose a question on whether 
and what kind of sophistication is required 
in order to obtain an I-V characteristics.
For a case of small diameter nanotubes, we note that a 
contribution to current by crossing subbands 
in metallic nanotubes under a moderate bias does not depend 
significantly on the  potential profile. 
The reason is the density of states of crossing subbands 
is nearly constant around the Fermi energy and therefore the 
transmission is insensitive to the changes in the potential profile. 
It is also clear that 
the renormalization of 
subbands due to  scattering 
affects only band edges, but does not influence 
the density of states 
of crossing subbands near the Fermi 
energy. 
These two facts, allow us to conclude that neither self-consistency
nor Kramers-Kr\"onig relation [Eq.~(\ref{eq:kram})] is necessary 
to obtain an I-V characteristics  in small diameter metallic nanotubes 
under a moderate bias. The criterium for this approximation is 
that bias is lower than the renormalized
bandgap of the first non crossing subband $E_{NC1}$, given by 
\begin{eqnarray}
E_{NC1} &=&   2 (|t_o||1 - 2 cos(\frac{\pi}{3} - \frac{\pi}{N})| 
 -\Delta E_{rn}) \label{eq:G0} \\
 &\sim &  2 (|t_o|\pi\sqrt{3}/N  -\Delta E_{rn}), \mbox{ for large }N.
 \nonumber  
\end{eqnarray}
Such an approximation, applicable {\it e.g.} to  a $(12,0)$ nanotube under 
a bias smaller than $1$V, while giving incorrect potential profiles would still
result in a correct  current with or without scattering.

\begin{figure}
\includegraphics [width=9.0cm] {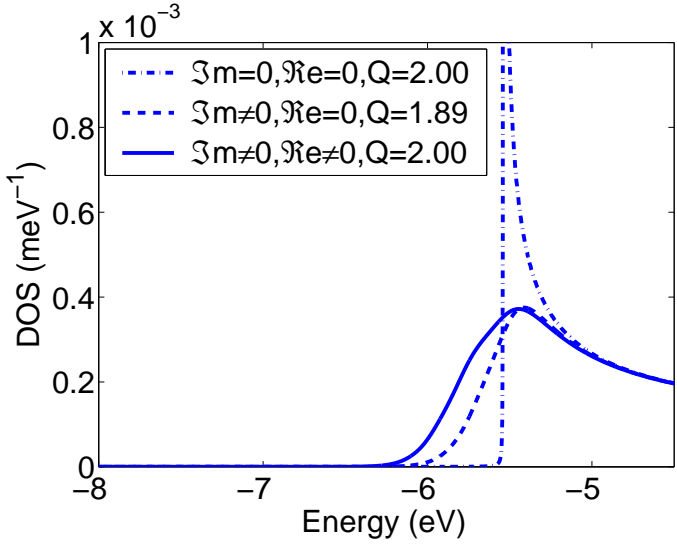}
\caption{Density of states in a $(12,0)$ nanotube under a zero bias,
for a fixed potential $V(y)=0$. 
The area under each curve $Q= \int DOS(E)dE$ 
gives twice the charge per subband.
Including only the imaginary part of 
scattering self-energy (dashed) results in a loss of charge.}
\label{fig:dos}
\end{figure}

\begin{figure}
\includegraphics [width=9.0cm] {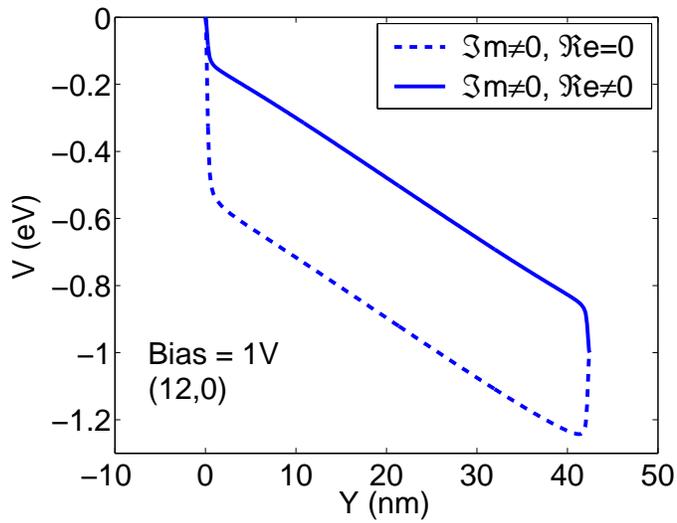}
\caption{Converged potential profiles versus position for (12,0) nanotube
for the cases when the real part of electron-phonon self-energy is neglected
(dashed) or taken into account (solid). Neglecting the real 
part results in a severe down shift of potential profile.}
\label{fig:potnore}
\end{figure}

As will be seen later, in studying current through large diameter nanotubes
it is important to take into account a current 
contribution by non crossing subbands.
Tunneling current by non crossing subbands has 
a lower threshold  bias [Eq.~(\ref{eq:G0})] and depends exponentially on
the slope of the potential near the edges of the tube.
Therefore, a self-consistent solution for the shape 
of tunneling barrier is must  for large diameter 
nanotubes at all biases. This, in turn, necessitates the exact knowledge
of electron charge and  density of states. 
We now discuss how much the real 
part of electron-phonon retarded self-energy $\Re e[\Sigma_{ph,q}^{R}(E)]$
and thus the renormalization of
the density of states  affects  the potential profile. 
In previous studies \cite{TED}  $\Re e[\Sigma_{ph,q}^{R}(E)]$ 
was set to zero, which naturally alleviates computational requirements.
In Fig.\ref{fig:dos} we show the density of states of crossing subbands 
versus  energy in a $(12,0)$
nanotube under a zero bias. The Poisson 
iterations were switched off and the potential $V(y)=0$.
Three curves represent 
different approximations to electron transport:
 a ballistic case, when both real and imaginary 
parts of electron-phonon self-energies are set to zero 
(dash-dotted line), 
a scattering case, when only imaginary part is taken 
into account (dashed line), 
and  a scattering case, when both imaginary and real part 
are non zero (solid line). For scattering cases, 
transport equations were iterated for long enough to achieve 
convergence of electron-phonon self-energies, {\it i.e.} self-consistent
Born approximation is satisfied.
The area under the ballistic curve $Q= \int DOS(E)dE$ is equal to $2$
which is twice the charge per subband. 
The important consequence of taking into account
only imaginary part of retarded self-energy is that the area $Q$ decreases,
resulting in $\sim5\%$ loss of electron charge.  When both real and 
imaginary parts are taken into account, electron charge is 
recovered due to the shift of the subband bottom.
The loss of charge when the real part is neglected results in a 
completely incorrect potential profile
when solving transport equations self-consistently with
Poisson equation [Eq.~(\ref{eq:prof})].
In Fig.\ref{fig:potnore} we show
potential profiles with and without a real part of 
electron-phonon self-energy when both Born and 
Poisson iterations have converged. 
The applied voltage drops symmetrically across the nanotube, when
the scattering is treated properly (solid line).
When the real part is neglected, the profile shows a severe down shift 
due to a missing electron charge (dashed line). Such 
incorrect potential profile will result in large error in current
due to non crossing subbands in large diameter nanotubes.
Another source of error is an overestimated 
bandgap [Eq.~(\ref{eq:G0})] and higher threshold for the onset of 
current due to non crossing subbands.
In this work, all our 
self-consistent calculations included Eq.~(\ref{eq:kram}) for the real 
part of electron-phonon self-energy.

\section{\label{sec:results} Results}
\subsection{\label{sec:low} Electrostatics at low bias}
\begin{figure}
\includegraphics [width=9.0cm] {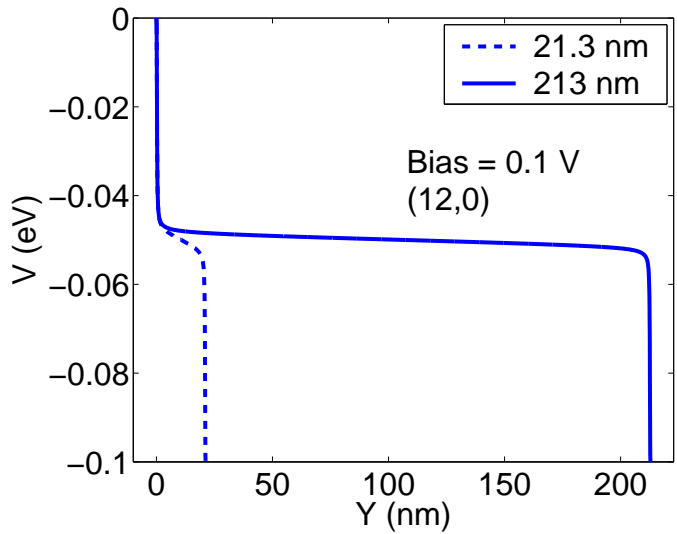}
\caption{Potential versus position for (12,0) nanotubes
of lengths 21.3 and 213 nm. The diameter of the (12,0) nanotube
is 0.94 nm and the applied bias is 100 mV.}
\label{fig:pot1}
\end{figure}

The mean free path due to scattering with acoustic phonons 
is in the range of a micron~\cite{yao-prl-00,park-nl-04}.
At low
biases, {\it i.e.} biases lower than inelastic phonon energy, 
electron-phonon scattering does not play a significant
role in determining the potential profile for wires of moderate length
(less than few hundreds of nanometer). The potential profiles for 
(12,0) nanotubes of 
lengths 21.3 and 213 nm are shown in Fig. \ref{fig:pot1}.
The edges of the nanotube near the contact rapidly screen the applied
bias / electric field. 
The potential drop is divided unequally
between different parts of the nanotube,
 with 90\% of the applied bias falling within 
1 nm from the edges for both lengths. 

\begin{figure}
\includegraphics [width=9.0cm] {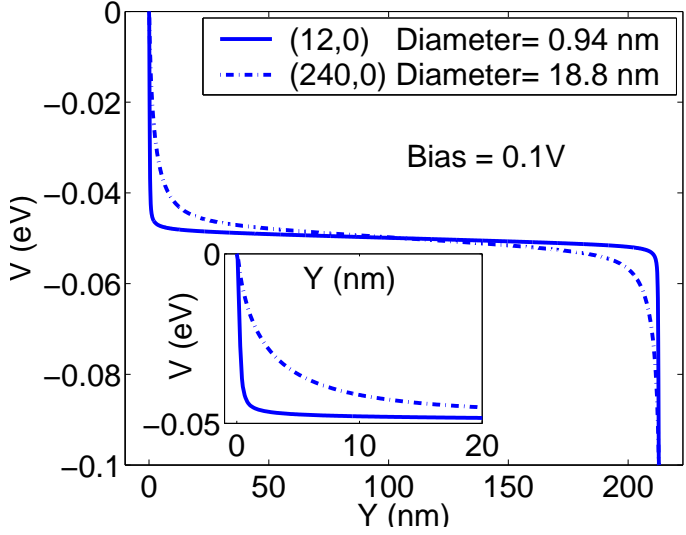}
\caption{Potential versus position for (12,0) and (240,0) nanotubes, 
which have diameters of 0.94 and 18.8 nm respectively. The screening for the large
diameter nanotube is significantly poorer. The inset magnifies the
potential close to the nanotube-contact interface. }
\label{fig:pot2}
\end{figure}

While the density of states (DOS) per unit length of
metallic nanotubes is independent of diameter, the DOS per atom is
inversely proportional to the diameter:
\begin{eqnarray}
DOS (E_F) / atom = \frac{2}{t_o N}~, \label{eq:dos}
\end{eqnarray}
where $N$ is the number of atoms in a ring of an (N,0) nanotube. 
As a result, we find that the
screening of metallic nanotubes degrades with diameter. The potential
drop for two nanotubes with diameters of 0.94 nm [(12,0) nanotube]
and 18.8 nm [(240,0) nanotube] are shown in Fig. \ref{fig:pot2}.
Clearly, screening is poorer in the larger diameter nanotube. In fact,
while the potential drops by 45 mV in a distance of 1 nm from the edge
for the (12,0) nanotube, the potential drop is only 17 mV for the 
(240,0) nanotube. The inset of Fig. \ref{fig:pot2} shows a 
substantially larger electric field away from the edges of the large
diameter nanotube. 

\begin{figure}
\includegraphics [width=9.0cm] {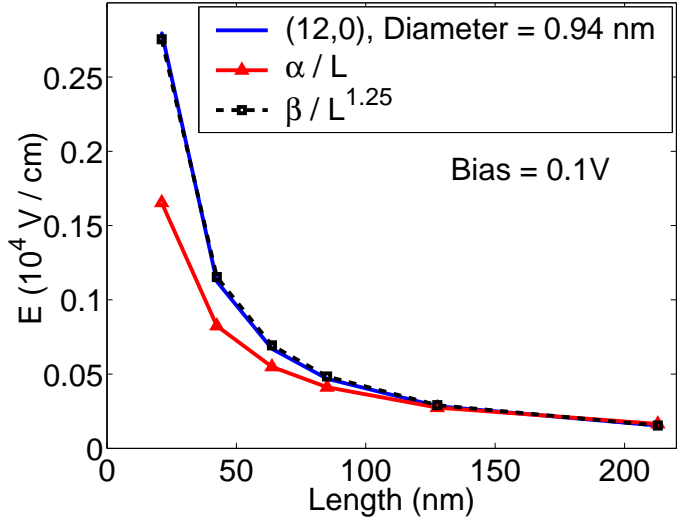}
\caption{
This plot shows the electric field at the mid point versus 
nanotube length. A (12,0) nanotube at an applied bias of 100 mV is 
considered. The electric field decreases approximately as 
$L^{-1.25}$.}
\label{fig:efield}
\end{figure}

The electric field at the center of the nanotube as a function of 
length is shown in Fig. \ref{fig:efield} for the tube with a diameter of
0.94 nm. We find that for all diameters, the electric field decreases
more rapidly than $\frac{1}{L}$, where $L$ is the length of the
nanotube. The exact power law however depends on the diameter.
If the computed electric field versus length is fit to $\frac{1}{L^a}$,
the exponent $a$ increases with increase in chirality. The value of
$a$ increases from 1.25 to 1.75 as the diameter increases from 0.94
to 18.85 nm. Similarly, electric field versus diameter
can be fit to $D^b$, where $b$ is in the range between 0 and 1.

\subsection{\label{sec:high} Electrostatics at high bias and 
current-voltage characteristics}
\begin{figure}
\includegraphics [width=9.0cm] {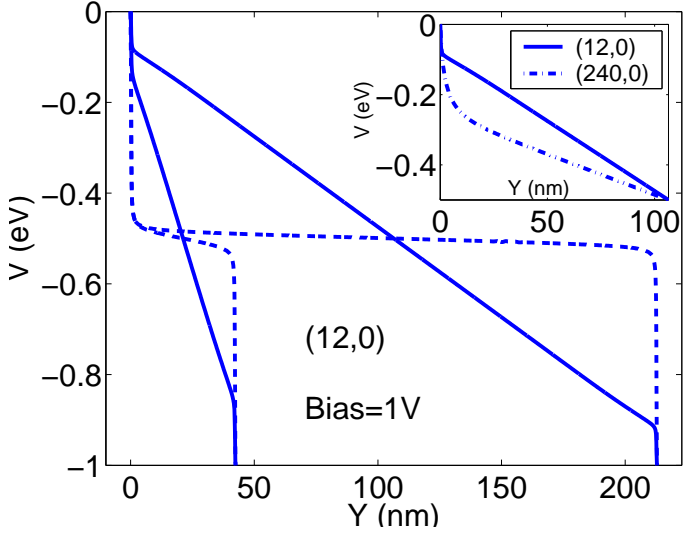}
\caption{The potential as a function of position is shown for (12,0) 
nanotubes of lengths 42.6 and 213 nm (solid). 
Shown for comparison are the
potential profile in the ballistic case (dashed). Inset: Potential versus 
position for two different diameters. Note that contrary to the low 
bias case,  the electric fields
away from the edges is larger for low diameter tube.
 The length of the nanotubes are both 213 nm.}
\label{fig:pot-hi-bias}
\end{figure}

At biases larger than 150 mV, the main scattering mechanism in 
defect free carbon nanotubes is electron-phonon interaction. 
Kane {\it et al.}\cite{yao-prl-00} found that emission of zone 
boundary and optical phonons are the dominant scattering 
mechanisms. The
potential profiles at high bias for 42.6 and 213 nm long (12,0) nanotubes are
shown in Fig. \ref{fig:pot-hi-bias}. 
Due to the increased resistivity of the tube 
the potential profile drops almost uniformly across
the entire nanotube, which is qualitatively different 
compared to the low bias
(and no scattering) results of Fig. \ref{fig:pot1}.
The potential drop at the edges of the nanotube accounts 
for only 30\% of the applied bias.
Because a bias drops in the bulk of the tube at the expense of 
the edges, the potential drop at the edges also decreases with 
increase of nanotube length, falling to 15\% for the longer nanotube. 
Due to the diameter dependence of the scattering rates,
the (12,0) tube is 20 times more resistive than (240,0), 
with mean free path of 5 nm for (12,0) versus 100 nm for (240,0).
As a result, contrary to the low bias case,
the potential profiles for larger  diameter tubes show
a larger voltage drop at the edges 
(inset of Fig. \ref{fig:pot-hi-bias}).

\begin{figure}
\includegraphics [width=9.0cm] {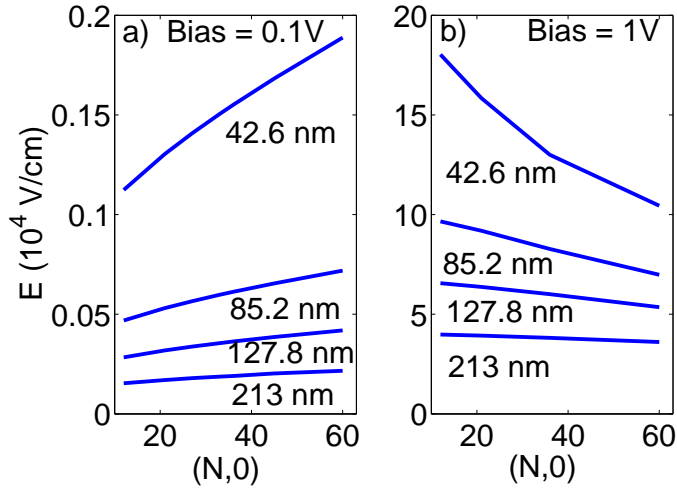}
\caption{Electric field as a function of diameter 
for nanotubes of lengths 42.6, 85.2, 127.8 and 213 nm. 
(a) At low bias, higher density of states in lower 
diameter tubes leads to the better screening (lower field). 
(b) At high bias, the trend is reversed and the electric 
field decreases with nanotube diameter.} 
\label{fig:pot-EfldvsN_LovsHibias}
\end{figure}

A major consequence of different potential profiles
of large and small diameter nanotubes at low and high bias 
is the electric field at the nanotube center:
the electric field is smaller in the small diameter nanotube in 
the ballistic limit but the onset of electron-phonon scattering at high bias
makes the electric field larger. 
This interesting reversal in electric field, demonstrated in 
Fig. \ref{fig:pot-EfldvsN_LovsHibias} is
rationalized by noting that the potential drop in the
(12,0) nanotube has become almost linear because the 
mean free path is much smaller than nanotube length, 
unlike in the (240,0) nanotube.

\begin{figure}
\includegraphics [width=9.0cm] {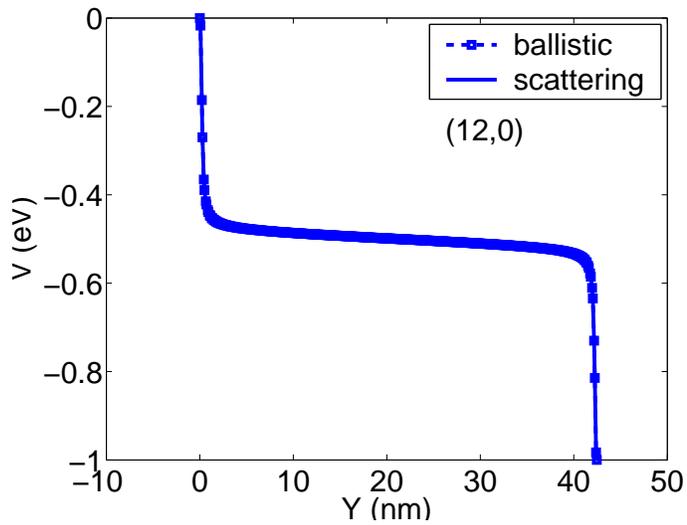}
\caption{Potential profile of a nanotube at 
equilibrium but in the presence of an electric 
field corresponding to a bias of 1V. 
Fermi energy is $-0.5$V  throughout the tube.
In contrast to the non equilibrium case 
shown in Fig.\ref{fig:pot-hi-bias}, at 
equilibrium the potentials with and without 
scattering agree with each other.}
\label{fig:pot-eq}
\end{figure}

So far electrostatics of nanotubes under applied bias was discussed.
 We now compare the 
potential profiles in the ballistic limit and with scattering, 
in a non zero electric field but at equilibrium. 
We choose an  external electric field corresponding to a $1$V bias but the 
Fermi levels of the source and drain contacts are set to $-0.5$V. 
The electrostatic potential profiles in the two cases are almost 
identical as shown in Fig.\ref{fig:pot-eq} in sharp contrast to 
the non equilibrium case. 
The reason is, at equilibrium, following 
the Thomas-Fermi model, the potential profile should depend only 
on DOS at Fermi energy, which is unaffected by electron-phonon 
scattering.

\begin{figure}
\includegraphics [width=9.0cm]  {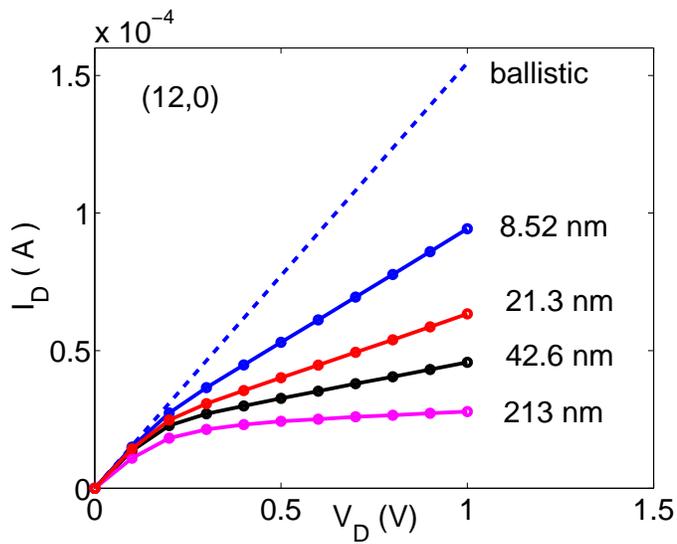}
\caption{Current voltage characteristics of a (12,0) 
nanotube of different lengths. The dashed line shows the ballistic limit.
The resistivity of the tube increases with 
the length causing current saturation.  
}
\label{fig:IVvsL}
\end{figure}

We now discuss the current voltage characteristics of small and large diameter nanotubes. 
For the (12,0) nanotube, only two crossing subbands contribute to transport.
The ballistic current increases linearly
with applied bias and the differential conductance is $4e^2/h$. 
Scattering by inelastic phonons causes current saturation and the decrease of the 
differential conductance (Fig. \ref{fig:IVvsL}). 
As the length of the nanotube 
is increased, so is the number of scattering events. 
The family of current - voltage 
characteristics show the transition 
from ballistic to diffusive transport regime.  
The current saturation at the value of 25 $\mu$A for the longest tube
agrees well with experimental data.

\begin{figure}
\begin{center}
\includegraphics [width=9.0cm]  {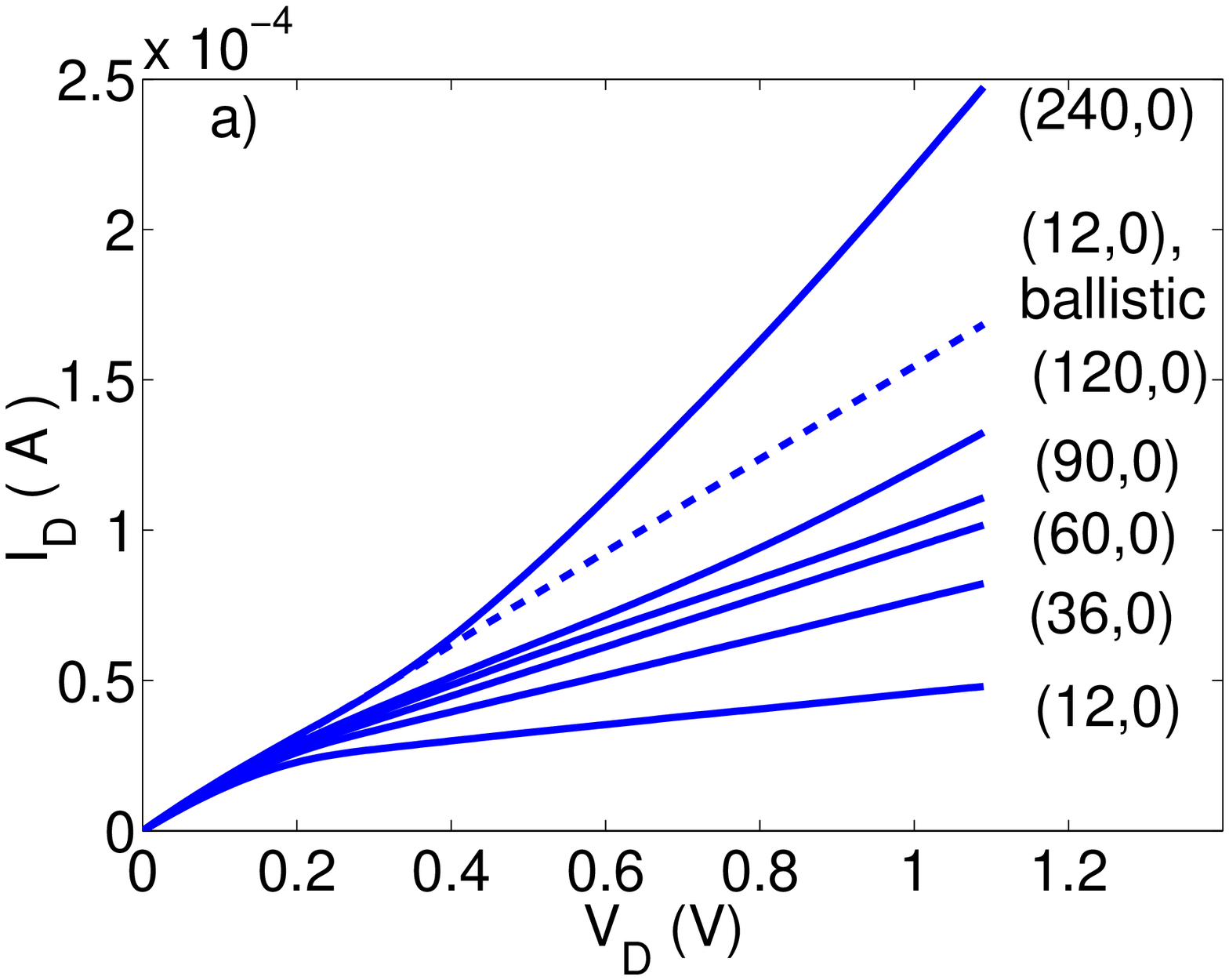}
\includegraphics [width=9.0cm]  {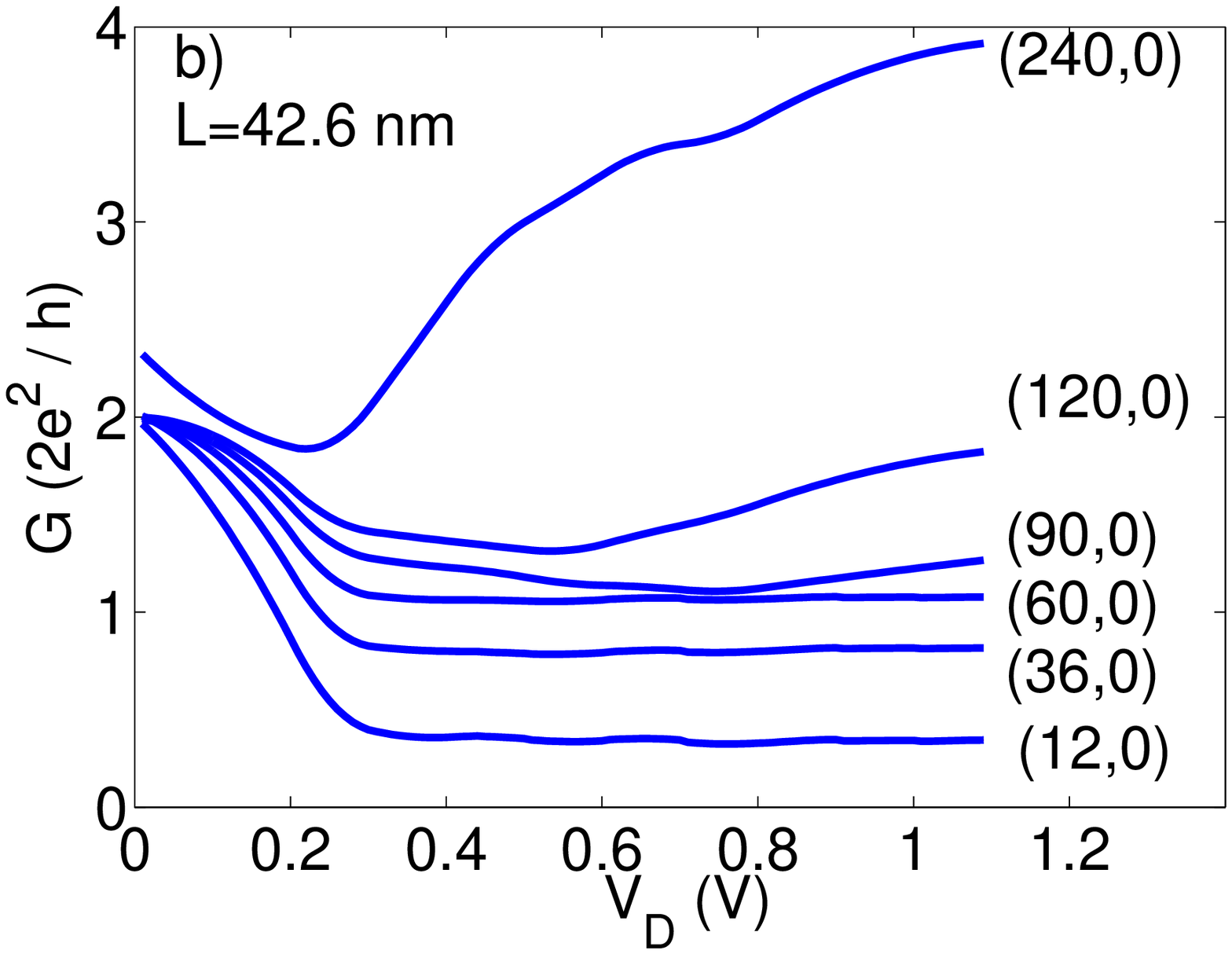}
\end{center}
\caption{(a) Current-voltage characteristics of nanotubes of 
various diameters, 
in the presence of electron-phonon scattering. (b) The differential 
conductance versus bias corresponding to (a). 
Inelastic phonon emission causes the decrease in differential
 conductance at high bias. Contribution of non crossing subbands 
 leads to the increased low and high bias conductance for large
  diameter tubes.
}
\label{fig:dIdV}
\end{figure}

Current and differential conductance versus bias for a $42.6$ nm nanotubes 
and a wide range of 
diameters are shown in 
Fig.\ref{fig:dIdV}.
At low bias, phonon scattering  is weak, so the current
and conductance are still close to the ballistic limit.
The saturation of current  corresponds to 
the onset of inelastic phonon scattering
and to the decrease of conductance at high bias. 
Differential conductance for small and
moderate diameter   nanotubes  ($N =$ 12, 36 and 60) is qualitatively 
similar because of the same number of conducting modes. Quantitatively, 
the conductance increases with diameter due to decreasing scattering 
rates.
This transition from low to high bias regime is also present 
in large diameter nanotubes ($N =$ 90, 120 and 240).

\begin{figure}
\includegraphics [width=8.0cm] {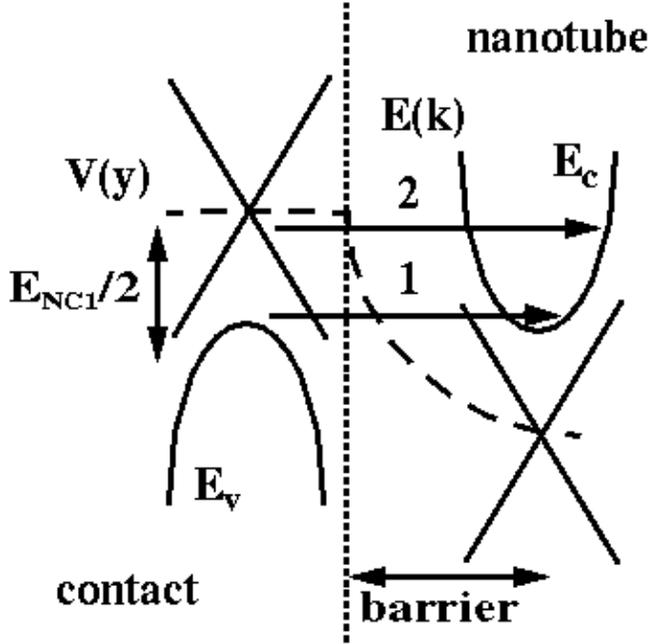}
\caption{Schematics for tunneling into non crossing subbands 
at the source. The 
dashed line shows a potential profile $V(y)$ and the dotted line shows 
a boundary between a contact and a nanotube. The arrow 1 shows 
injection from a perfect contact with a bandstructure of a nanotube.
The arrow 2 shows injection from end-contacts 
with constant density of states or a scattering-assisted tunneling
from  crossing into non crossing subbands.
}
\label{fig:zener}
\end{figure}

We note that  current increases  with diameter. 
In the ballistic limit, the self-consistently
calculated current of (240,0) nanotube  at 1V is 310 $\mu$A
and the differential conductance 
is almost $13 e^2 / h$. When scattering is included the current
is decreased to 218.3 $\mu$A which is much larger than the 
current carried by a (12,0) nanotube of the same length, which is
45.7 $\mu$A. In addition, 
the differential conductance versus bias exhibits a 
qualitative change in shape. It is bell-shaped for small diameter 
nanotubes and transforms to U-shaped for large diameter nanotubes.
As explained in \cite{svizhenko-cnt-ball},
the increase of differential conductance with bias 
occurs due to {\it injection of electrons from contacts 
into the low energy  non crossing subbands}. 
A schematics of Zener tunneling process is shown in Fig.\ref{fig:zener}:
 when the bias 
becomes larger than twice the bandgap of the lowest non crossing subband,
electrons can tunnel from  valence band 
states ($E<-E_{NC1}/2+V(y)$) in the source to  
the conduction band  states ($E >E_{NC1}/2+V(y)$) in the 
middle of the tube (the channel) and also
from  valence band 
states in the channel  to the conduction band  states in the drain.
The lowest non crossing subbands ($q=N/3-1$) in (90,0), (120,0) and (240,0) 
in the ballistic limit have bandgaps $E_{NC1}=$ 331, 249 and 125 meV 
respectively 
and start to contribute 
to current at biases of twice these values. 
 We note 
that the above picture is based on the assumption that the 
applied voltage drops symmetrically 
across the nanotube. In the case when 
one contact is much more resistive than the other and the applied
voltage drops completely at one of the edges, the threshold for Zener tunneling
is reduced, starting at biases equal to $E_{NC1}$ rather than $2E_{NC1}$. 
Additionally, if the contacts are metallic (and not perfect nanotube 
contacts), then the threshold  bias for tunneling into non crossing 
subbands will be further reduced by a factor of 2, an issue that will be discussed in 
section \ref{sec:contacts}. In addition to Zener tunneling between states with the same 
quantum number $q$, scattering also 
induces a phonon assisted tunneling  from  bonding states of non 
crossing subbands to crossing subbands. Although these processes 
require intersubband scattering, the threshold bias is twice as low.

Another important feature of differential conductance 
occurs at zero bias in (240,0) 
nanotube (Fig.\ref{fig:dIdV}).
The zero bias conductance of (240,0) nanotube
is larger than $4e^2/h$ because the 
non crossing subbands  are partially 
filled and contribute to current:
the first non crossing subband opens at an energy of 2.4 $kT$ 
from the band center.
This contribution is a simple intraband transport, 
determined by the population of the conduction band in the source and the 
valence band in the drain, but rather insensitive to 
the details of the potential profile.
At slightly higher biases
the non crossing subbands  contribution  to current 
saturates to a constant value and the contribution to differential conductance decreases to zero, while
the total conductance decreases to $4e^2/h$.

Increasing the length of a nanotube eventually makes the 
nanotube length larger than the mean free path. The current carried 
by a 213 nm long (240,0) nanotube is now  86.6 $\mu$A at 1V.
Note 
that the differential conductance of  a long nanotube, 
shown in the inset of Fig.\ref{fig:POT_MWNT},
has changed to bell-shaped, in contrast to the short nanotube case 
presented in Fig.\ref{fig:dIdV}. This is because when the nanotube 
length is many times the mean free path, the potential drop in 
the nanotube bulk becomes more linear and as a result the barrier 
width (Fig.\ref{fig:zener}) for tunneling into non crossing 
subbands increases. The, non crossing subbands then do not carry 
significant current. Over all, the voltage drop at the edges is 
a crucial factor determining the current contribution due to non 
crossing subbands. It is directly related to the height and width 
of the barrier for tunneling into non crossing subbands 
(Fig.\ref{fig:zener}), which depends on the scattering rate, 
nanotube length and screening capability. 
 
In the rest of the paper we discuss the role of electrostatic coupling 
to inner shells of a large diameter nanotube and the role of contacts. 
Specifically, we will see that both issues lead to increased tunneling 
into non crossing subbands because they lead to a decrease in barrier 
thickness.

\subsection{\label{sec:shells}Electrostatic gating due to inner shells}
\begin{figure}
\includegraphics [width=9.0cm] {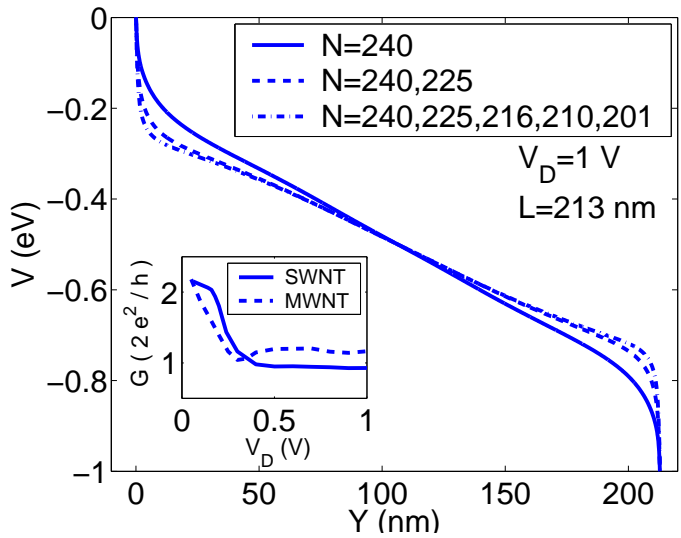}
\caption{The potential profile in a single- and multiwalled NT. 
The barrier width in the outer (240,0) shell 
decreases due to the additional 
screening provided by the inner shells. Inset: a comparison of differential 
conductance in singlewalled (240,0) nanotube (dashed) and multiwalled 
nanotube of the same diameter. Both curves are bell-shaped, signifying
a quenching of Zener tunneling.  
}
\label{fig:POT_MWNT}
\end{figure}

In multiwalled nanotubes (MWNT) current flows only through the 
outermost shell as the inner shells are not electronically 
coupled to the metal leads \cite{liang-apl-04}, and inter 
shell hopping is negligible \cite{bourlon-prl-04}. Here, we are 
interested in the role of electrostatic coupling between 
the inner and outer shells, in affecting the potential profile 
and current carried by the outer shell. We find that the inner 
shells effectively decrease the barrier width for tunneling into 
non crossing subbands by providing additional screening. 
Fig.\ref{fig:POT_MWNT} shows the potential profile 
in the outermost (240,0) shell. The inner shells are 
chosen to be metallic nanotubes with chirality indices 
$N =$ 225, 216, 210 and 201, such that a separation between 
the walls is roughly $3~$\AA. Even when just one metallic 
shell (225,0) is added, the barrier width for tunneling into 
the first non crossing subband decreases dramatically from 2.97 
to 0.97 nm. Adding more shells further reduces the barrier width 
by smaller amounts, and the effect saturates. The differential 
conductance with the inner shells is qualitatively different as 
shown for the 213 nm long (240,0) nanotube in the inset of 
Fig.\ref{fig:POT_MWNT}. We find that the differential 
conductance at higher biases is larger but does not show an 
increase with bias, when inner shells are included.

\subsection{\label{sec:contacts} Influence of contact quality}
We have so far assumed that the contacts are perfect, {\it i.e.} 
made of semi-infinite carbon nanotube leads. As mentioned in the 
introduction, this physically corresponds to carbon nanotubes 
weakly coupled to the metal in which they are buried. So, 
injection of electrons into the carbon nanotube lying between 
the contacts occurs from the carbon nanotube buried in the metal. 
As a result of this, injection into non crossing subbands from 
the contact cannot occur at E=0. 
Note that selection rules don't 
permit injection of electrons from the crossing subband of the 
nanotube contact into non crossing subbands between the contacts. 
In this section, we relax this condition and consider injection 
from the source contact into all nanotube subbands. The drain 
contact continues to be a nanotube contact. To accomplish this 
in a phenomenological manner, we model the contact self 
energies in Eq.~(\ref{eq:surf_sigma}) as,
\begin{eqnarray}
\Sigma^R_{S,q} = -i \alpha \rho t_q^2 \mbox{ ,} \label{eq:self-gold}
\end{eqnarray}
where $t_q$ is the original hopping parameter in Eq.~(\ref{eq:surf_sigma}), 
$\alpha$ is called the quality factor of the contacts, and 
$\rho$ is the density of states of the contact which is chosen 
to be close to that of gold, $\rho = 0.17 eV^{-1}$. The self 
energy in Eq.~(\ref{eq:self-gold}) permits injection into the 
first non crossing subband at a bias of $E_{NC1}$ rather than 
the $2E_{NC1}$ with nanotube contacts.

\begin{figure}
\includegraphics [width=9.0cm] {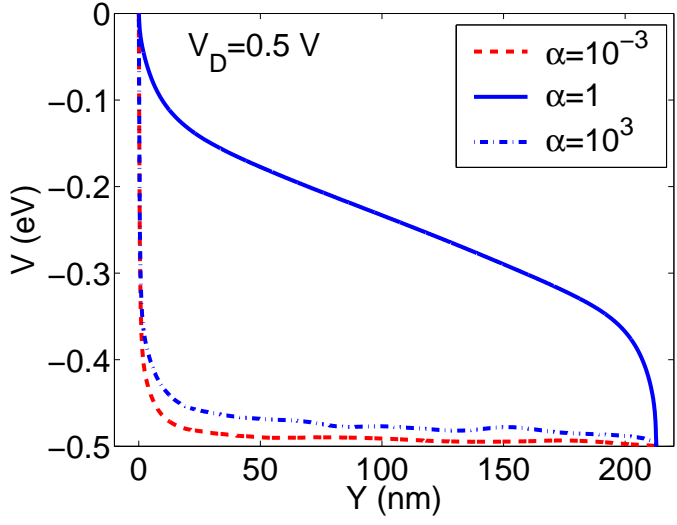}
\caption{Potential profile for good ($\alpha=1$) and poor 
($\alpha=10^{-3}$ and $\alpha=10^{3}$)  source contacts. 
When the coupling is poor, voltage drops mostly at the source, 
which opens Zener tunneling for higher subbands.
}
\label{fig:POTvsYMS}
\end{figure}

\begin{figure}
\includegraphics [width=9.0cm] {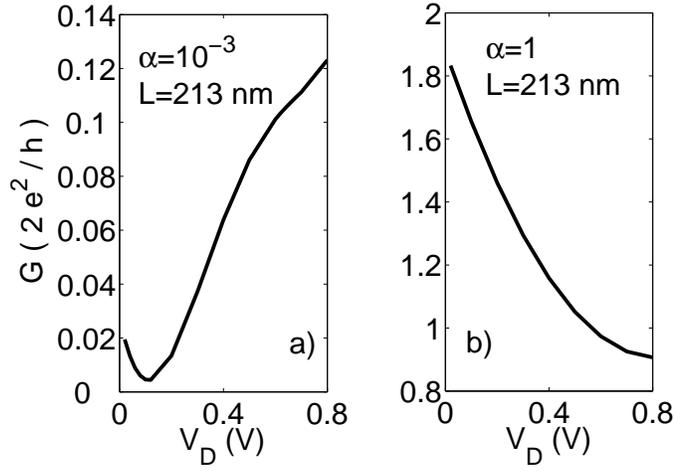}
\caption{Differential conductance  $G$ vs. voltage $V_D$ for a) poor 
($\alpha=10^{-3}$) and b) good  ($\alpha=1$) source contacts. 
Presence of Zener tunneling at $\alpha=10^{-3}$
qualitatively changes G-V characteristics.
}
\label{fig:GVMS}
\end{figure}

\begin{figure}
\includegraphics [width=9.0cm] {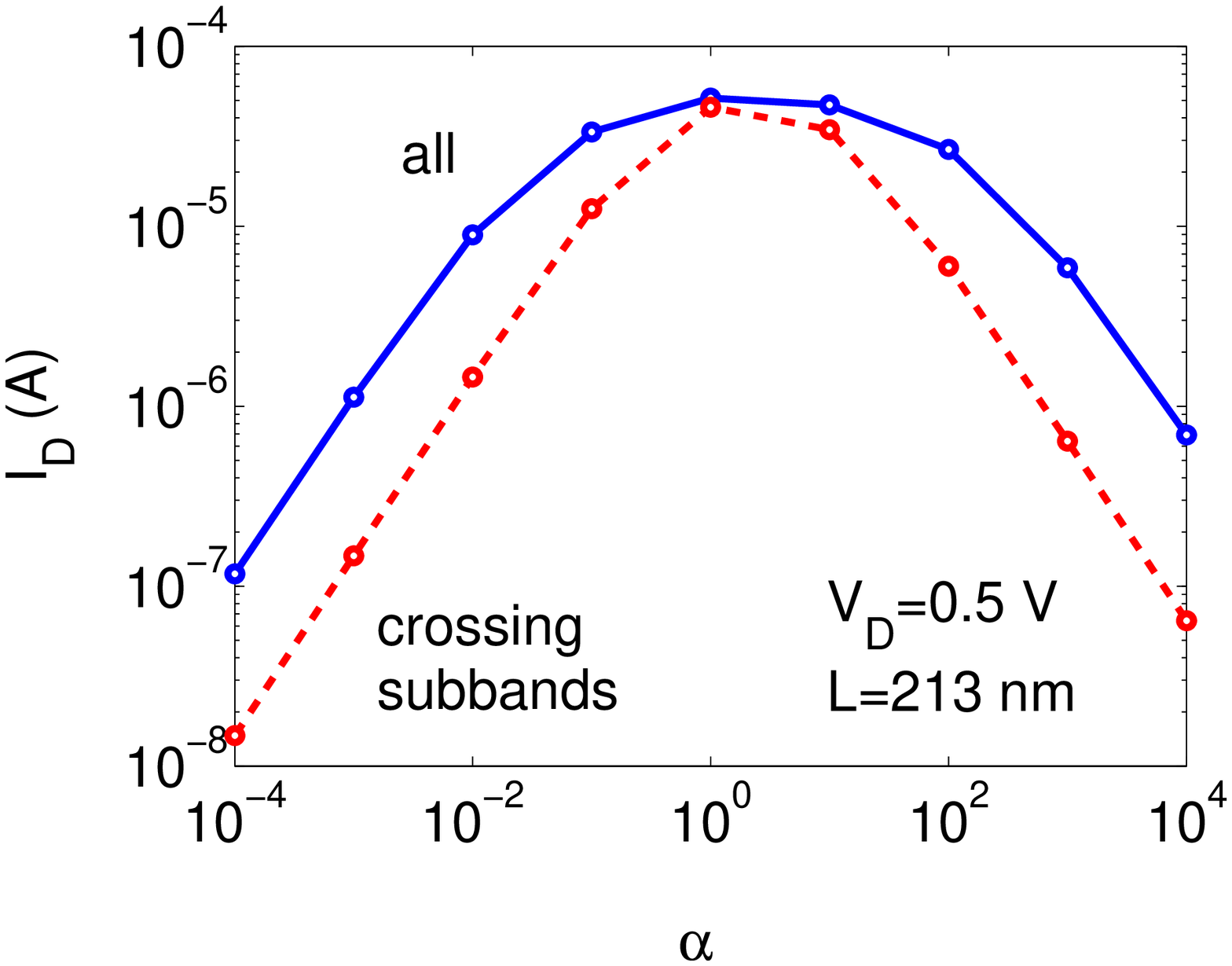}
\caption{Current as a function of a source contact quality factor  $\alpha$.
When the contact is poor (far from $1$),
the effect of higher subbands becomes important which signifies
the increasing trend of differential conductance versus bias.
}
\label{fig:IvsCoup}
\end{figure}

We now present results for the potential profiles with good 
($\alpha=1$) and poor ($\alpha=10^{-3}$ and $\alpha=10^3$) contacts 
in Fig.\ref{fig:POTvsYMS}. Note that the terminology of poor 
and good contact is a relative to the intrinsic resistance of 
the nanotube that arises due to electron-phonon scattering. 
For $\alpha=1$, the potential profile of a (240,0) nanotube 
of length 213 nm is more or less symmetric and qualitatively 
similar to the perfect contact results presented in 
Fig.\ref{fig:POT_MWNT}. For $\alpha=10^{-3}$ and $\alpha=10^3$, 
the electrostatic potential drops predominantly at the 
source-end due to the large source-end contact resistance. 
The large potential drop and hence extremely thin barrier in the source-end 
will facilitate tunneling into the non crossing subbands for 
poor contact. As mentioned before, the asymmetry of the potential profile further 
reduces the threshold for Zener tunneling to biases of $E_{NC1}/2$. 
The difference between the potential profiles 
for good and poor contacts has a profound influence on the 
shape of the differential conductance versus bias as shown 
in Fig.\ref{fig:GVMS}. For good contacts, the 213 nm long 
nanotube shows a decreasing conductance with bias, in agreement 
with the perfect contact case. In contrast to this, for poor 
contacts, the differential conductance increases with bias. 
This increase with bias is a direct consequence of the 
potential profile shown by the dashed line in 
Fig.\ref{fig:POTvsYMS}, which facilitates significant 
injection into non crossing subbands. To see more directly 
that the non crossing subbands are important in carrying 
current in long nanotubes with poor contacts, we calculate 
the current carried at a drain bias of 0.5 V as a function 
of the quality factor of the contacts, with only crossing 
subbands and with all subbands.  For good contacts, the 
current with only crossing subbands is very close to the 
current with all subbands (Fig.\ref{fig:IvsCoup}). But for 
poor contacts, the current with all subbands is significantly
 higher than the current with only crossing subbands. For example,
  when $\alpha=10^{-3}$, the current with all subbands is nearly 
  an order of magnitude larger than with only crossing subbands. 
From Fig.\ref{fig:IvsCoup} one can also see that 
the U-shaped differential conductance curve versus bias should be observed
in a wide range of $\alpha$ except 
a small region near $\alpha\sim 1$, where non 
crossing subbands do not carry a significant current.

\section{\label{sec:conc} Conclusion}
We have studied transport in nanotubes of varying diameters and 
length, with electron-phonon scattering included. We find that 
charge self-consistency and the proper treatment of 
subband renormalization due to scattering  are
 crucial in determining the correct 
current-voltage characteristics in large diameter metallic 
nanotubes. In the small bias ballistic limit, while the applied
 bias drops predominantly at the nanotube-contact interfaces, 
 screening is incomplete at the tube center. Further, screening 
 improves with decrease in nanotube diameter due to the increased 
 density of states per atom near the Fermi energy. At biases 
 larger than 150 mV, electron-phonon scattering becomes 
 important and the electrostatic potential drops primarily 
 in the bulk of the nanotube rather than at nanotube-contact
  interfaces. However, as the mean free path for electron-phonon 
  scattering increases with increase in nanotube diameter, the 
  potential drop in the bulk of the nanotube is larger for small 
  diameter nanotubes. As a result, the electric field at the 
  nanotube center increases with increase in diameter at small
   biases, and decreases with increase in diameter at high 
   biases. This interesting reversal in electric field versus 
   diameter is computationally seen for nanotube lengths from 
   42 to 213 nm. 
Over all, we find that larger diameter nanotubes are capable 
of carrying more current because of increase in mean free path 
with increase in diameter and lower bias threshold for injection 
into non crossing subbands.
For small diameter nanotubes of length 42 nm, we find that the 
differential conductance versus bias is bell-shaped, with the 
largest value at zero bias. The reason for the smaller differential
 conductance at larger bias is reflection of electrons due to
  inelastic phonon scattering and availability of only crossing 
  subbands for charge transport. In contrast, large diameter 
  nanotubes of the same length show an increase in differential 
  conductance with increase in bias due to injection into non 
  crossing subbands and larger mean free paths. At a nanotube 
  length (213 nm) longer than the mean free path, we find that 
  the differential conductance for large diameter nanotubes 
  transitions to a bell-shaped curve similar to the small 
  diameter case, when coupling to contacts is good (small 
  contact resistance). 
We have also modeled the role of 
       inner shells in affecting the potential profile of 
       the outer shell of a multiwalled nanotube. Here, we find
        that the inner shells cause a change in the electrostatic 
        potential profile of the outer shell so as to make the
         potential drop sharper at the nanotube-contact interfaces.
          Our computational study has shown that the potential 
          drop and differential conductance in metallic nanotubes
           is determined by an interesting interplay of diameter,
            mean free path, nature of contacts and bias threshold 
            for tunneling into non crossing subbands. 
 Finally, the most important result of this paper 
 is that  {\it a quality of contacts is a primary factor in 
      determining the 
       shape of the differential conductance versus bias in 
      large diameter nanotubes}. When 
   the resistance at the nanotube-contact interface is larger
    than the intrinsic resistance of the nanotube, there is a
     large potential drop at the interface. This potential drop 
     facilitates considerable tunneling into non crossing subbands 
     and as result the differential conductance of the 213 nm 
     long (240,0) nanotube with poor contact increases with 
     bias, in contrast to the case with good contacts. 
      This finding gives a possible explanation to 
       the increase in differential conductance 
       with bias seen in the recent experiment of reference 
       \cite{liang-apl-04}.

\begin{acknowledgments}
The calculations were performed at the facilities of NASA Advanced 
Supercomputing Division. We would like to thank Avik Ghosh for
 discussions on the density of states of gold.
\end{acknowledgments}

\pagebreak



\begin{thebibliography}{1}


\bibitem{frank-science-98}
S. Frank,
P. Poncharal, Z. L. Wang and W. A. de Heer,
Science {\bf 280}, 1744 (1998)

\bibitem{poncharal-epjd-99}
P. Poncharal, S. Frank, Z. L. Wang and W. A. de Heer,
Eur. Phys. J. D {\bf 9}, 77 (1999)

\bibitem{pablo-apl-99}
P. J. de Pablo,
E. Graugnard, B. Walsh, R. P. Andres, S. Datta and R. Reifenberger,
Appl. Phys. Lett. {\bf 74}, 323 (1999)

\bibitem{nygard-applphysa-99}
J. Nygard, D. H. Cobden, M. Bockrath, P. L. McEuen and P. E. Lindelof,
Appl. Phys. A {\bf 69}, 297-304 (1999)

\bibitem{kong-prl-01}
J. Kong, E. Yenilmez, T. W. Tombler, W. Kim, H. Dai, R. B. Laughlin,
L. Liu, C. S. Jayanthi, and S. Y. Wu,
Phys. Rev. Lett. {\bf 87}, 106801 (2001)


\bibitem{kreupl-microeleceng-02}
F. Kreupl, A. P. Graham, G. S. Duesberg, W. Steinhogl, M. Liebau, E. Unger
and W. Honlein,
Microelectron. Eng. {\bf 64}, 399-408 (2002)

\bibitem{ngo-eeetn-04}
Q. Ngo, D. Petranovic, S. Krishnan, A. M. Cassell, Q. Ye, J. Li,
M. Meyyappan and C. Y. Yang,
IEEE Trans. Nanotech. {\bf 3}, 311 (2004)

\bibitem{wei-apl-01}
B. Q. Wei, R. Vajtai and P. M. Ajayan,
Appl. Phys. Lett. {\bf 79}, 1172-1174 (2001)


\bibitem{yao-prl-00}
Z. Yao, C. L. Kane and C. Dekker,
Phys. Rev. Lett., {\bf 84}, 2941 (2000)

\bibitem{collins-prl-01}
P. G. Collins, M. Hersam, M. Arnold, R.Martel and Ph. Avouris,
Phys. Rev. Lett. {\bf 86}, 3128-3131 (2001)

\bibitem{pablo-prl-02}
P. J. de Pablo, C. Gomez-Navarro, J. Colchero, P. A. Serena,
J. Gomez-Herrero and A. M. Baro,
Phys. Rev. Lett.,  {\bf 88}, 36804 (2002)

\bibitem{poncharal-jpcb-02}
P. Poncharal, C. Berger, Y. Yi, Z. L. Wang and W. A. de Heer,
J. Phys. Chem. B {\bf 106}, 12104 (2002)

\bibitem{liang-apl-04}
Y. X. Liang, Q. H. Li and T. H. Wang,
Appl. Phys. Lett. {\bf 84}, 3379-3381 (2004)



\bibitem{Anantram-prb-00}
M. P. Anantram,
Phys. Rev. B {\bf 62}, 4837 (2000)

\bibitem{svizhenko-cnt-ball}
A. Svizhenko, M. P. Anantram, T.R. Govindan,
submitted to {\it IEEE Trans. on Nano Tech.}, (2005)



\bibitem{tans} S.J. Tans, M. Devoret, H. Dai, A. Thess, 
R.E. Smalley, L.J. Geerligs,
and C. Dekker, Nature, v. 386, p. 474 (1997)


\bibitem{Anantram-apl-01}
M. P. Anantram, 
Appl. Phys. Lett., v. 78, p. 2055 (2001)


\bibitem{JAP}
A. Svizhenko, M. P. Anantram, T.R. Govindan, R. Venugopal,
J. Appl. Phys., {\bf 91}, 2343 (2002)

\bibitem{park-nl-04}
J. Park, S. Rosenblatt, Y. Yaish, V. Sazonova, H. Ustunel, S. Braig, T. A.
Arias, P. W. Brouwer and P. L. McEuen,
Nano Lett. {\bf 4}, 517 (2004)

\bibitem{golgsman-prb-03}
G. Pennington and N. Goldsman,
electrons is semiconducting carbon nanotubes
Phys. Rev. B {\bf 68}, 045426 (2003) 


\bibitem{TED}
A. Svizhenko, M. P. Anantram,
IEEE Ttrans. Electron. Dev., {\bf 50}, 1459 (2003)



\bibitem{bourlon-prl-04}
B. Bourlon, C. Miko, L. Forro, D. C. Glattli, A. Bachtold,
Phys. Rev. Lett., {\bf 93}, 176806 (2004) 





\end{thebibliography}
\end{document}